\documentclass{jfm}
\usepackage{graphicx}
\usepackage{epstopdf, epsfig}
\usepackage{xcolor}
\usepackage[export]{adjustbox}
\usepackage{float}
\usepackage{amsmath}
\usepackage[normalem]{ulem}
\usepackage{subcaption}

\def\We{{\it We}}
\def\Oh{{\it Oh}}
\def\Bo{{\it Bo}}
\def\AR{{\it AR}}

\def\u{{\bf u}}
\def\j{{\bf j}}

\shorttitle{Partial coalescence dynamics}
\shortauthor{M.R. Behera, H. Deka, K. C. Sahu and G. Biswas}

\title{Effect of velocity, fluid properties and drop shape on coalescence and neck oscillation}

\author{Manas Ranjan Behera\aff{1},
\ Hiranya Deka\aff{2},
\ Kirti Chandra Sahu\aff{3},
\ and Gautam Biswas\aff{4,5}
 \corresp{\email{gtm@iitk.ac.in, gautamb@goa.bits-pilani.ac.in}}}

\affiliation{\aff{1} 
Department of Chemical Engineering, BITS Pilani, K. K. Birla Goa Campus, Goa 403726, India
\aff{2}Department of Mechanical, Materials and Aerospace Engineering, Indian Institute of Technology Dharwad, Dharwad 580007, India
\aff{3}Department of Chemical Engineering, Indian Institute of Technology Hyderabad, Kandi 502284, Telangana, India
\aff{4}  Department of Mechanical Engineering, Indian Institute of Technology Kanpur, Kanpur 208016, India
\aff{5} Department of Mechanical Engineering, BITS Pilani, K. K. Birla Goa Campus, Goa 403726, India}

\begin{document}

\maketitle
\begin{abstract}
We perform axisymmetric numerical simulations to investigate the coalescence dynamics of a liquid drop in a deep liquid pool. This study aims to generalize the mechanisms of partial coalescence across a range of drop shapes, elucidate the underlying mechanism of neck oscillations, and examine the roles of inertial, viscous and gravitational forces, quantified by the Weber, Ohnesorge, and Bond numbers, in governing the coalescence behavior. A phase diagram is constructed to delineate the boundaries between partial and complete coalescence regimes based on these dimensionless parameters. Our analysis of the height-to-neck ratio shows that, upon contact with the pool, the primary drop forms an upward liquid column that ultimately pinches off due to inwardly directed horizontal momentum. Additionally, the study suggests that as the dimensionless numbers increase, the effect of the vertical collapse rate plays a significant role in the outcome of the coalescence process. Notably, the Rayleigh-Plateau instability is found to be insignificant in driving partial coalescence within the explored parameter space. We identified a transition regime between partial and complete coalescence, characterized by multiple neck oscillations that delay the pinch-off of secondary droplets. The formation of secondary droplets is most prominent for prolate drops, followed by spherical and oblate drops of comparable volume. Furthermore, we observe that the tendency to form multiple droplets from elongated liquid columns diminishes with an increase in the impact velocity of the primary drop.
\end{abstract}


\section{Introduction}\label{sec:intro}

The coalescence of droplets on liquid surfaces has garnered significant attention due to its importance in a wide range of applications, including atomization and spray processes, oil–water separation in enhanced oil recovery, coatings, microfluidics, combustion, and inkjet printing \citep{stone2004engineering, thoroddsen2008high, kavehpour2015coalescence}. It also plays a crucial role in natural phenomena such as the impacts of raindrops on water bodies, the ejection of marine microplastics by rainfall, soil erosion \citep{low1982collision, veron2015ocean,liu2018experimental}, and gas exchange driven by air capture at the air-water interface  \citep{thomson1886v,pumphrey1989underwater}. Beyond its practical relevance, droplet coalescence is of fundamental interest \citep{bouwhuis2012maximal,wang2013we,kirar2022coalescence} as it involves rich multiscale interfacial dynamics and air-film drainage processes \citep{tran2013air,hendrix2016universal,deka2019coalescence}.

When a liquid drop impacts a liquid pool, the resulting dynamics is governed by a complex interplay of inertia, capillarity, viscosity, gravity, and thin-film effects. Depending on the relative importance of these forces, impact outcomes are commonly classified into floating \citep{Couder2005c}, bouncing \citep{alventosa2023inertio}, coalescence \citep{thoroddsen2000pof, blanchette2006partial}, and splashing \citep{worthington1908study,ray2015regimes,dandekar2025splash} regimes. These regimes encompass a wide range of phenomena, including regular bubble entrapment \citep{Oguz1990a}, vortex-ring formation \citep{behera2021}, vortex shedding \citep{thoraval2012, thoraval2013drop}, ejecta sheet formation \citep{agbaglah2015drop, fudge2023drop}, crown splashing \citep{paul2024splashing}, jetting \citep{das2022evolution}, and large-bubble entrapment \citep{ThoravalPRE, deka2017regime}. More recent investigations have focused on resolving the transition dynamics between these regimes \citep{behera2023investigation, sprittles2023gas}. The present work focuses on the coalescence regime, with particular emphasis on its underlying mechanisms and the transitions that emerge across a broad range of controlling parameters.

Coalescence dynamics involves the merging of two identical liquid bodies following the drainage and rupture of the thin fluid film trapped between the impacting drop and the liquid pool. Depending on the interfacial morphology during this process, two distinct outcomes may occur: partial coalescence, in which the drop pinches off to form a secondary daughter droplet, and complete coalescence, in which the drop merges entirely into the pool. The mechanisms underlying partial coalescence, as well as the conditions governing the transition from partial to complete coalescence, have attracted sustained scientific attention. \citet{Charles1960a,Charles1960b} were among the first to report partial coalescence in emulsion studies and delineated the boundary between partial and complete coalescence using the viscosity ratio, $\mu^* = \mu_l / \mu_a$, where $\mu_l$ is the viscosity of the drop/pool liquid and $\mu_a$ that of the surrounding fluid. They observed partial coalescence to occur predominantly within the range $0.02 < \mu^* < 11$. Subsequently, \citet{Fedorchenko2004} introduced the Ohnesorge number, $\Oh = \mu_l / \sqrt{\rho_l \sigma D_{eq}}$, to characterize drop coalescence behavior, capturing the balance among viscous, inertial, and capillary forces. Here, $\rho_l$ denotes the liquid density, $D_{eq}$ the equivalent spherical diameter of the drop, and $\sigma$ the interfacial tension between the drop and the surrounding medium. Considerable effort in the literature has focused on determining the critical Ohnesorge number $(Oh_c)$ that marks the transition from partial to complete coalescence. However, reported thresholds vary widely due to differences in the definitions of $Oh$ and the fact that most experiments were performed in the low-inertia regime. In particular, \citet{Gilet2007} and \citet{ray2010generation} distinguished between two Ohnesorge numbers: $Oh_1$, based on the viscosity of the drop/pool liquid, and $Oh_2$, based on the viscosity of the surrounding fluid, and consequently reported two critical values, $Oh_{1c}$ and $Oh_{2c}$.  \citet{Aryafar2006} combined data from air–liquid and liquid–liquid systems and defined a single critical Ohnesorge number based on the viscosity of the more viscous phase. While this unified representation facilitates comparison across diverse systems, it does not explicitly distinguish whether the transition from partial to complete coalescence is governed by the viscosity of the surrounding fluid or that of the drop/pool liquid. When results are presented in such a unified framework, as in \citet{Aryafar2006} and \citet{kavehpour2015coalescence}, a generalized criterion of $Oh_c \gtrsim 1$ emerges as a sufficient condition for complete coalescence across a wide range of systems. The present study, however, focuses on air–liquid interfaces, for which $Oh_2 \ll 1$. This enables isolation of the role of the liquid-phase viscosity, determination of $Oh_{1c}$, and a systematic investigation of the $Oh_1$-controlled transition from partial to complete coalescence. Although the Ohnesorge number determines whether viscous or inertia–capillary effects control the pinch-off process, gravitational forces also play an important role \citep{Chen2006a,blanchette2006partial,Gilet2007,ray2010generation,Behera2022}. The influence of gravity is quantified by the Bond number, $\Bo = \rho_l g D_{eq}^2/\sigma$, where $g$ denotes the gravitational acceleration; it measures the relative importance of gravitational to capillary forces. \citet{Chen2006b} reported that partial coalescence occurs primarily in the inertia–capillary regime, corresponding to intermediate values of $\Bo$ and $\Oh$. At high $\Bo$, gravitational drainage becomes dominant, thereby suppressing partial coalescence.

As discussed above, most previous studies have focused on the roles of capillarity, viscosity, and gravity in determining whether a drop undergoes partial or complete coalescence, typically under conditions where inertial effects are negligible. Consequently, the explicit and systematic influence of inertia has received comparatively little attention. Inertial effects can be quantified by the Weber number, $\We = \rho_l V^2 D_{eq} / \sigma$, which represents the ratio of inertial to capillary forces. The impact velocity $(V)$ is expected to influence the drainage of liquid from the drop into the pool and thereby affect the onset of partial coalescence, an aspect that warrants further investigation.

Although regime maps based on $\Oh$ and $\Bo$ provide valuable criteria for predicting the transition from partial to complete coalescence, the underlying physical mechanisms governing pinch-off and daughter-droplet formation remain actively debated. Early work by \citet{Charles1960b} attributed the pinch-off process to a Rayleigh–Plateau instability of the liquid column. In contrast, \citet{blanchette2006partial} demonstrated that partial coalescence is governed by the competition between the vertical and horizontal collapse rates of the primary drop, both driven by surface tension. This interpretation was subsequently supported by \citet{ray2010generation,Ding2012,Zhang2015b}. Other studies \citep{Honey2006,Chen2006a,Chen2006b} have also emphasized the role of capillary instabilities in partial coalescence. More recently, \citet{Angeli2023} revisited the problem and argued that Rayleigh–Plateau instability plays a crucial role in partial coalescence, with capillary waves facilitating the formation of the liquid column. These differing interpretations motivate the present study, which revisits the intricate dynamics of partial coalescence of an impacting drop, with particular emphasis on the role of inertia.

Another intriguing aspect is the post-coalescence behavior of the liquid column formed after impact. In some situations, this column does not pinch off immediately but instead undergoes neck oscillations, resulting in the formation of multiple secondary structures. Although this behavior has been noted in prior studies \citep{Zhang2009, ray2010generation, Alhareth2020}, the underlying mechanisms, especially the influence of impact velocity and drop shape, remain elusive. Furthermore, a falling droplet undergoing oblate–spherical–prolate shape oscillations \citep{ThoravalPRE, balla2019shape, deka2019dynamics, agrawal2020experimental} can also affect the coalescence process during its interaction with a liquid pool, presenting another intriguing area for exploration \citep{anirudh2024coalescence, deka2019dynamics, zhangb2019}. \citet{thoraval2013drop} recognized that shape oscillations in water droplets with diameters exceeding the capillary length scale, $\sqrt{\sigma/(\rho_l g)}$, can affect both the coalescence and impact dynamics on a liquid layer. This phenomenon has also been experimentally investigated by \citet{dighe2024extremely}, who found that such larger drops tend to flatten during descent, resulting in shallower cavity formation upon impact. The cavity formation becomes more pronounced at higher velocities. Additionally, oblate drops with large aspect ratios exhibit asymmetric coalescence \citep{anirudh2024coalescence} and display intriguing splashing dynamics \citep{anirudh2026unveiling}. Upon contacting the free surface over a broad area, the parent drop displaces air, forming an annular region that traps a ring-shaped bubble. This trapped air leads to fragmentation into smaller bubbles that rise asymmetrically. A capillary wave then generates a deep crater, triggering intense whipping action in the later stages of coalescence. The rapid ascent of the liquid column, driven by strong upward momentum, forms a tall structure that pinches off to produce one daughter droplet, while the remaining fluid in the column forms a second, non-spherical blob. In the context of drop impact on a solid substrate, \cite{yun2022role} demonstrated that ellipsoidal drops influence impact dynamics by inducing symmetry breaking in mass and momentum distribution, with oscillation phase and ellipticity playing crucial roles in modifying hydrodynamics and suppressing rebound. While considerable research has focused on spherical droplets, natural and engineered systems often involve nonspherical drops that undergo shape oscillations during free fall. These oscillations, manifesting as oblate–spherical–prolate deformations, can profoundly influence impact dynamics by altering the initial contact geometry, symmetry of coalescence, and the pathways of energy redistribution upon impact. Furthermore, the associated air entrapment, crater asymmetry, and fragmentation behavior introduce additional complexity, which remains inadequately understood \citep{anirudh2024coalescence, thoraval2013drop, dighe2024extremely}.

Motivated by these gaps, the present study systematically investigates the coalescence dynamics of both spherical and nonspherical droplets impacting a deep liquid pool with varying liquid properties, across a wide range of impact velocities and drop diameters, using axisymmetric numerical simulations. The primary objective is to unravel the underlying mechanisms governing partial coalescence and to identify the dominant forces driving the process. Special attention is given to the dynamics of neck oscillations, which are analyzed to gain deeper insights into the pinch-off phenomenon and the formation of secondary droplets. Overall, this study aims to develop a comprehensive understanding of the factors influencing droplet coalescence.

The remainder of the manuscript is organized as follows: the problem formulation, numerical methodology, and validation are described in \S\ref{sec:form}; the results and their discussion are presented in \S\ref{sec:dis}; and concluding remarks are provided in \S\ref{sec:conc}.
 
\section{Formulation} \label{sec:form}
\begin{figure}
\centering
\hspace{-0.5cm} (a) \hspace{5.5cm} (b) \\
\includegraphics[width=0.45\textwidth]{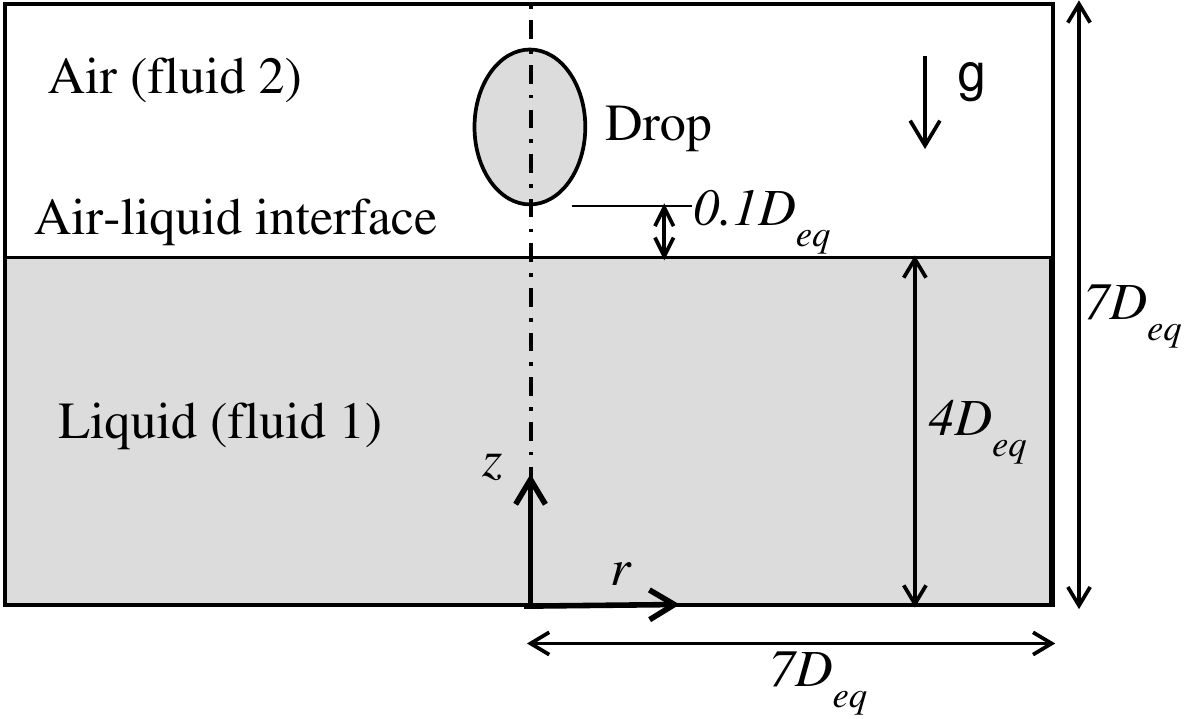} \includegraphics[width=0.45\textwidth]{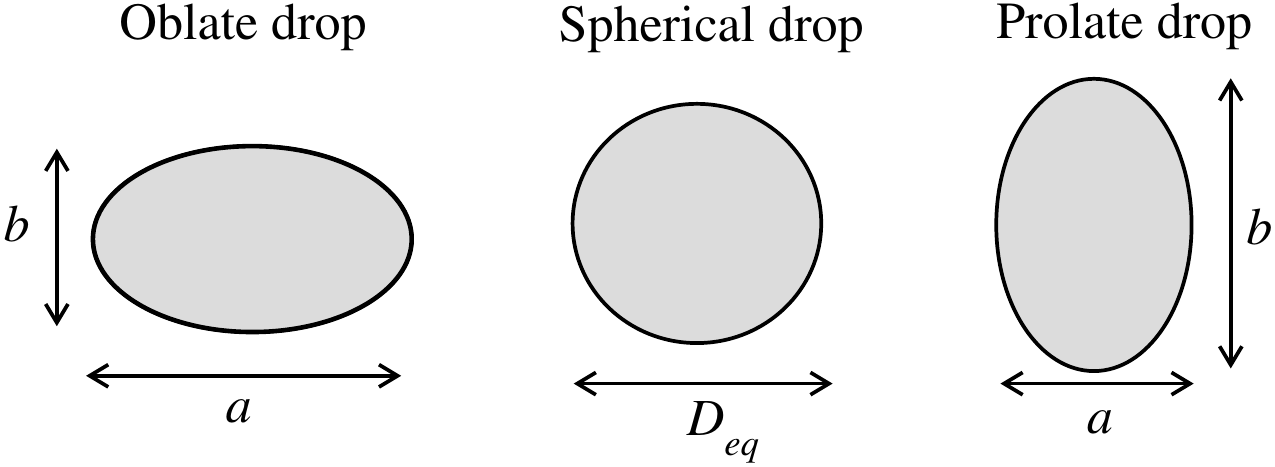}
\caption{(a) Schematic diagram illustrating the initial configuration of a drop impacting a liquid pool. (b) Depiction of various shapes of  drops. The aspect ratios, $\AR=b/a<1$, $\AR=1$, and $\AR>1$, correspond to oblate, spherical, and prolate shapes, respectively.}
\label{fig:1}
\end{figure}

We investigate the coalescence dynamics of a liquid droplet in a liquid pool using Gerris \citep{Popinet2003,Popinet2009}, an open source flow solver developed within the finite-volume framework. The flow dynamics is modeled using a cylindrical coordinate system $(r, z)$, with $z = 0$ denoting the bottom of the domain and $r=0$ representing the axis of symmetry. Figure \ref{fig:1}(a) illustrates a schematic diagram of the initial configuration of a drop that impacts a liquid pool. The initial shape of the drop is characterized by its aspect ratio, $\AR=b/a$, where $a$ and $b$ denote the horizontal and vertical diameters of the drop, respectively. Consequently, $\AR<1$,  $\AR=1$ and  $\AR>1$ signify oblate, spherical, and prolate shapes, respectively (see Figure \ref{fig:1}(b)). In the present study, we vary the initial aspect ratio $(\AR)$ while maintaining a constant volume, such that $D_{eq}^3=a^2b$, where $D_{eq}$ represents the volume-equivalent spherical diameter of the droplet. The liquids constituting the drop and the pool are denoted as fluid `1', while the surrounding medium (air) is designated as fluid `2'. While presenting the simulation results, red, green, and blue colours are employed to represent the drop liquid, pool liquid, and air, respectively. Both fluids are assumed to be Newtonian and incompressible. The dynamic viscosities and densities of fluid ‘1’ and fluid ‘2’ are $(\mu_l, \rho_l)$ and $(\mu_a, \rho_a)$, respectively.  Gravity $(g)$ acts in the negative $z$ direction. The size of the axisymmetric computational domain is $7D_{eq} \times 7D_{eq}$, with the initial height of the liquid pool set at $4D_{eq}$. We confirm that the computational domain size is big enough to ensure negligible boundary effects. At $t = 0$, a small gap of $0.1D_{eq}$ is maintained between the south pole of the drop and the free surface of the liquid pool, as shown in Figure \ref{fig:1}(a), with the drop having a prescribed initial velocity. 
%1

\subsection{Governing equations}

The governing equations, describing the dynamics of a liquid drop falling on a liquid pool with air as the surrounding medium, consist of the equations of mass and momentum conservation. The dimensional form of these governing equations is given by:
\begin{equation}
\rho \left[ \frac{\partial \u}{\partial t} + \u \cdot \nabla \u \right] = -\nabla p + \nabla \cdot \left[\mu(\nabla \u + \nabla \u^T)\right] + \sigma \kappa \delta_s \hat{n} - \rho g \j, \label{eq1}
\end{equation}
\begin{equation}
\nabla \cdot \u= 0. \label{eq2}
\end{equation}

Here, ${\bf u} = (u,v)$ represents the velocity field, where $u$ and $v$ denote the velocity components in the $r$ and $z$ directions, respectively. The pressure field is denoted by $p$, $t$ represents time, $\j$ stands for the unit vector in the $z$ direction, $\delta_s$ signifies the Dirac-delta function utilized to model a localized surface force at the interface, $\kappa$ denotes the curvature, and $n$ represents the outward-pointing (away from the liquid) unit normal to the interface.

The interface separating the air and liquid phases is tracked by solving an advection equation for the volume fraction of the liquid phase, denoted as $c$, where $c = 0$ and $1$ represent the air and liquid phases, respectively. This advection equation is expressed as:
\begin{equation} 
\frac{\partial c}{\partial t} + \u \cdot \nabla c = 0. \label{eq3}
\end{equation}
Thus, the density $(\rho)$, and the viscosity $(\mu)$ appearing in eq. (\ref{eq1}) depend on $c$ as
\begin{equation}
\rho = c \rho_l + (1 - c) \rho_a,
\end{equation}
\begin{equation}
\mu = c \mu_l + (1 - c) \mu_a.
\end{equation}
The governing equations (eqs. \ref{eq1}–\ref{eq3}) are solved with the following boundary conditions. At $r = 0$, a symmetry boundary condition is applied, while at the side boundary $(r = 7D_{eq})$, a free-slip boundary condition is employed. No-slip and no-penetration boundary conditions are utilized at the bottom wall $(z = 0)$. At the top of the computational domain ($z = 7D_{eq}$), a Neumann boundary condition is applied to velocity, ensuring a zero normal gradient for velocity (${\partial u/\partial z} = 0$), which represents a free-slip boundary. Additionally, a Dirichlet condition is imposed on pressure, with $P = 0$.

\subsection{Numerical method and validation}

%2
\begin{figure}
\centering
\hspace{0.6cm} (a) \hspace{5.7cm} (b)  \\ 
\includegraphics[width=0.45\textwidth]{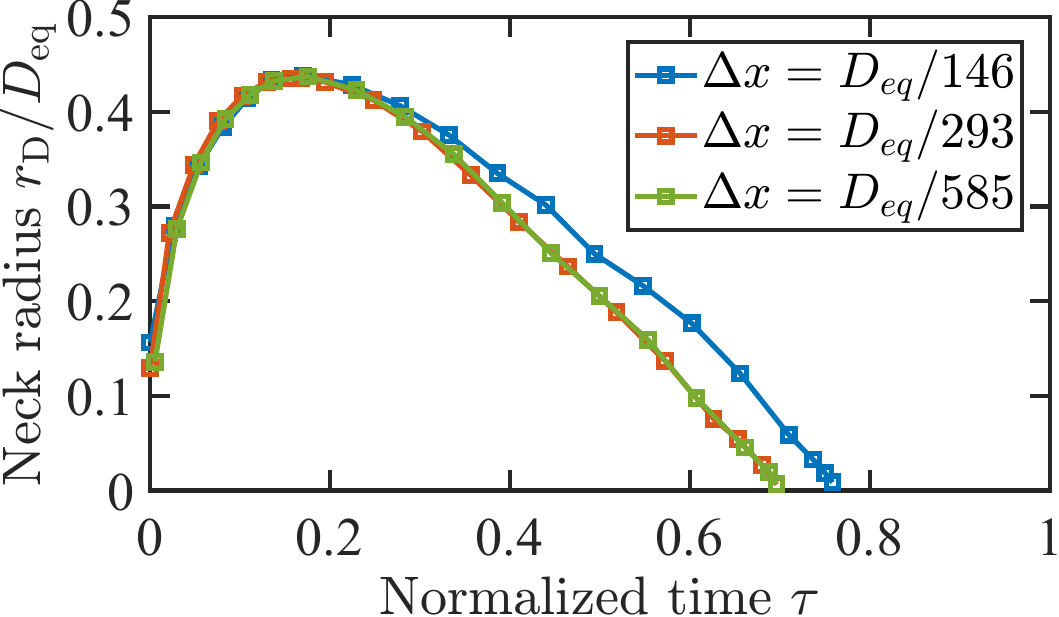} \hspace{2mm}
\includegraphics[width=0.45\textwidth]{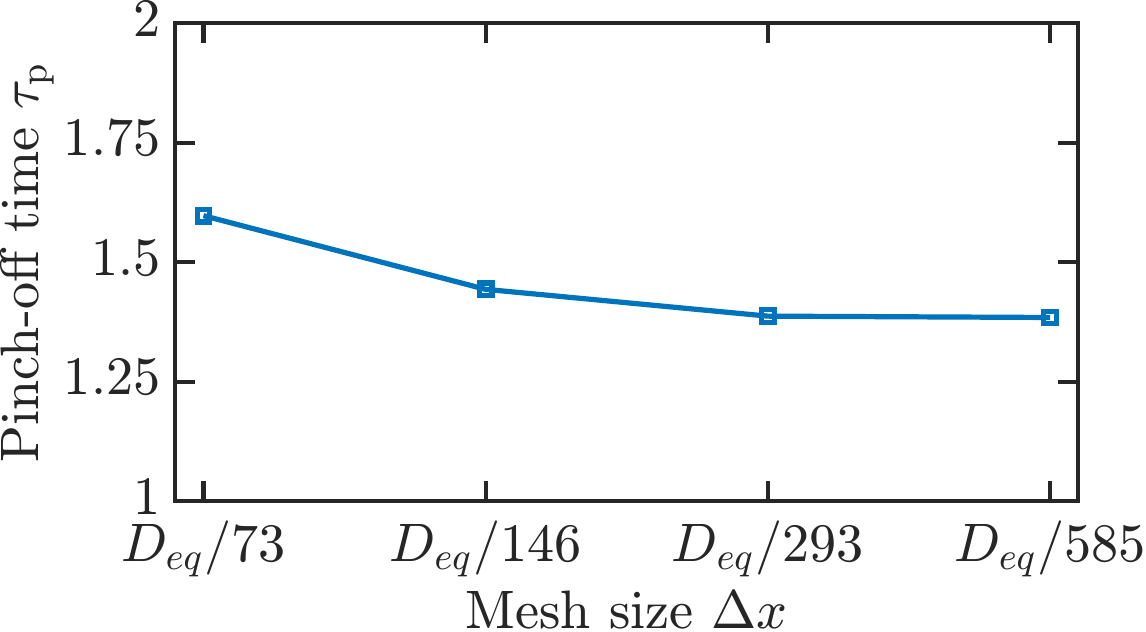} 
\caption{(a) Variation of the normalized neck radius $(r_D/D_{eq})$ with normalized time $(\tau = t/\sqrt{\rho_l D_{eq}^3/8\sigma})$ for different mesh sizes $\Delta x$, with $\AR = 1.4$, $\We = 40$, $\Bo = 0.5$, and $\Oh = 0.003$. (b) Variation of normalized pinch-off time $(\tau_{p} = t_{p}/\sqrt{\rho_l D_{eq}^3/8\sigma})$ with mesh size $\Delta x$ for $\AR = 1$, $\We = 0$, $\Bo = 0.5$, and $\Oh = 0.014$.}
\label{fig_grid}
\end{figure}

The open-source flow solver Gerris \citep{Popinet2003,Popinet2009}, employed in this study, uses a Volume-of-Fluid (VoF) method coupled with dynamic adaptive mesh refinement based on vorticity magnitude and proximity to the interface. The flow solver has been deployed earlier by \cite{thoraval_thesis} to investigate drop impact, splashing and air entrapment, entailing a stellar contribution pertaining to understanding of gas-layer and interfacial dynamics. Gerris implements a numerical scheme that is second-order accurate in both space and time and incorporates a height-function-based balanced force formulation to include surface tension in the Navier–Stokes equations (Eq.~\ref{eq1}). The two-dimensional computational domain is discretized into $(2^n)^2$ cells, where $n$ denotes the refinement level, allowing for adaptive adjustment of cell sizes between the minimum and maximum refinement limits. The adaptive refinement strategy employed in this study incorporates several features to accurately capture interfacial dynamics while maintaining computational efficiency accurately: (i) interface tracking using gradient thresholds of the VoF function $c$ to resolve phase boundaries, (ii) vorticity-based refinement to capture rotational flow features, and (iii) region-specific geometric criteria to refine areas with high-velocity gradients or thin-film behavior. This approach effectively resolves key interfacial phenomena such as air film rupture, neck formation, and the propagation of capillary waves along the drop surface. To validate the numerical approach, we performed a grid convergence study for a representative parameter set. Figure~\ref{fig_grid}(a) shows the evolution of the normalized neck radius, $(r_D / D_{eq})$, as a function of the normalized time, $\tau = t / \sqrt{\rho_l D_{eq}^3 / 8\sigma}$, for different mesh sizes ($\Delta x = D_{eq}/146$ to $D_{eq}/585$). Notably, the results obtained with $\Delta x = D_{eq}/293$ and $\Delta x = D_{eq}/585$ are qualitatively indistinguishable. To enable a quantitative assessment, Figure~\ref{fig_grid}(b) presents the normalized pinch-off time ($\tau_{p}$) as a function of mesh size, revealing percentage differences of 9.6\% (between $\Delta x = D_{eq}/73$ and $D_{eq}/146$), 3.9\% (between $\Delta x = D_{eq}/146$ and $D_{eq}/293$), and only 0.2\% (between $\Delta x = D_{eq}/293$ and $D_{eq}/585$). In comparison to the larger deviations observed at coarser resolutions, the negligible difference between mesh sizes $\Delta x = D_{eq}/293$ and $D_{eq}/585$ confirms grid convergence. Accordingly, a mesh size of $\Delta x = D_{eq}/293$ is adopted for all simulations in this study, ensuring an optimal balance between numerical accuracy and computational cost.

%3
\begin{figure}
\centering
\includegraphics[width=0.95\textwidth]{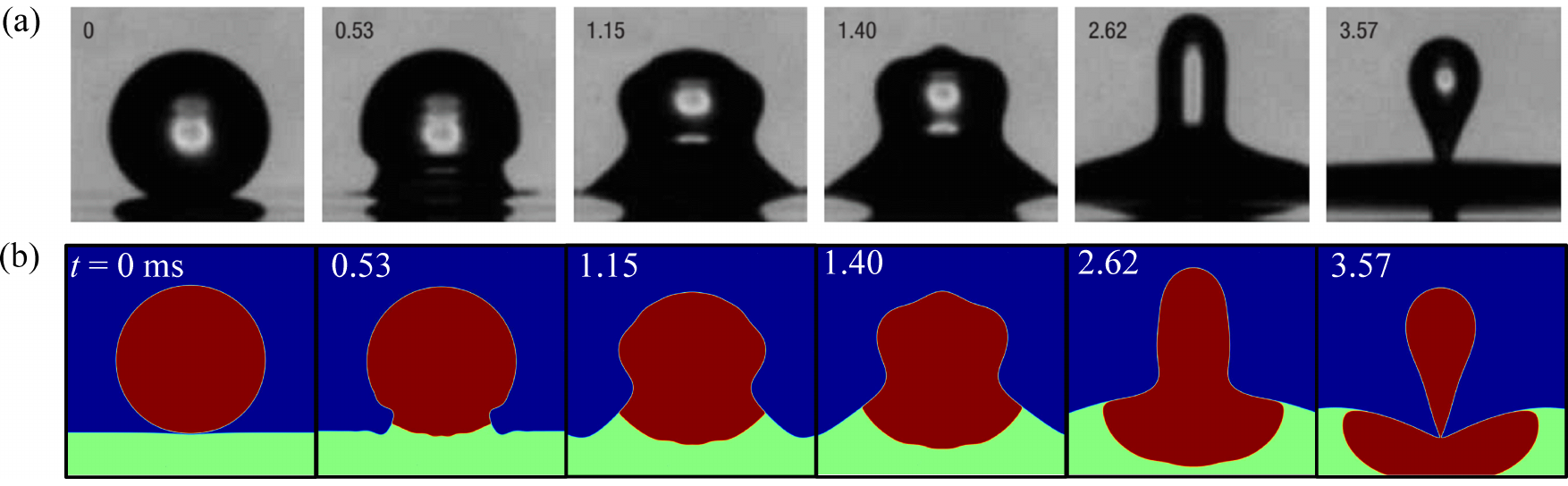}
\caption{The comparison between the results obtained using the current numerical solver and those of \cite{blanchette2006partial}. Experimental results from \cite{blanchette2006partial} are depicted in panel (a), while the results obtained in this study are displayed in panel (b). These panels illustrate the partial coalescence phenomenon of an ethanol drop of radius 0.535 mm in an ethanol pool with air as the surrounding medium. The times (in ms) are indicated on each image for reference.}
\label{fig_valid}
\end{figure}

%4
\begin{figure}
\centering
\includegraphics[width=0.85\textwidth]{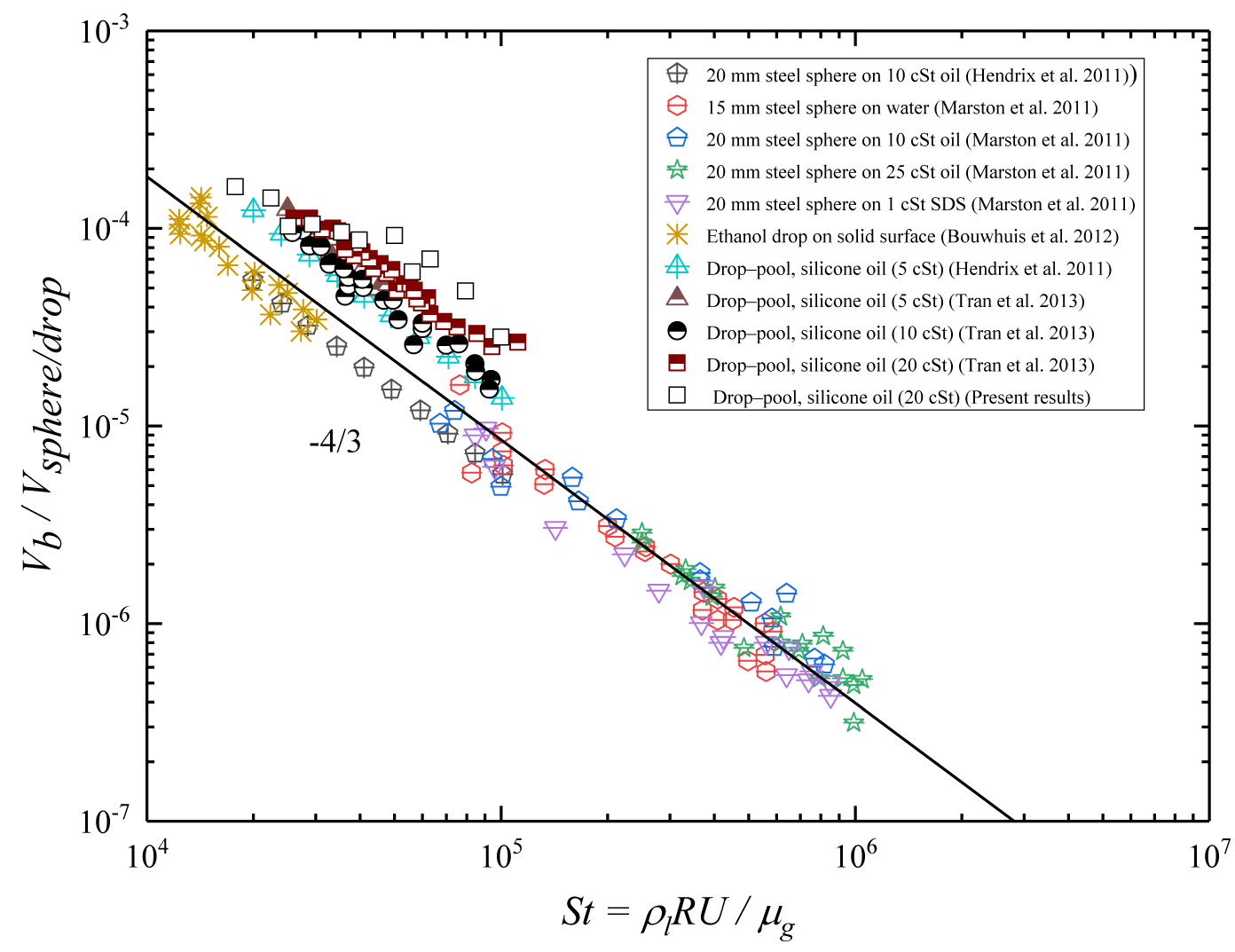}
\caption{Figure adapted from \citet{hendrix2016universal}. Log–log plot of the normalized bubble volume $(V_b / V_{sphere/drop})$ as a function of the Stokes number $(St)$. Here, $V_b$ and $V_{sphere/drop}$ are the volumes of the bubble and the impacting drop/sphere, respectively. The figure shows boundary-integral simulation results from \citet{hendrix2016universal}, together with previously published experimental data from \citet{tran2013air}, \citet{Marston_2011}, and \citet{bouwhuis2012maximal}. Results from the present study are superimposed as open squares. The solid line denotes the $-4/3$ scaling law originally reported by \citet{tran2013air}.}
\label{fig_validb}
\end{figure}

The numerical solver used in the present study is available in the following GitHub repository (https://github.com/manas27526/Impact-dynamics-of-liquid-drops). Several researchers have also extensively utilized Gerris to explore interfacial flows, and it has been validated for various configurations  \citep{deka2019coalescence,behera2023investigation,ThoravalPRE,agrawal2017nonspherical}. Additionally, in Figure \ref{fig_valid}, we compare the coalescence dynamics of an ethanol drop (of radius 0.535 mm) obtained from our present numerical simulation with the experiment conducted by \cite{blanchette2006partial}. Here, $t = 0$ represents the instant at the onset of the coalescence process. As time progresses ($t > 0$), the drop begins to drain into the pool, as observed at $t = 0.53$, 1.15 and 1.4 ms. This drainage leads to the formation of a cylindrical column with a height exceeding that of the initial drop at $t = 2.62$ ms. Subsequently, the neck of this column narrows due to the inward pull of surface tension, resulting in the pinching off of a secondary drop at $t = 3.57$ ms. The comparison reveals that the dynamics captured by our numerical simulation at various times agree well with the experimental observations of \cite{blanchette2006partial}. Furthermore, Figure~\ref{fig_validb} compares the entrapped-bubble volumes from the present numerical simulations of drop impact on a liquid pool with available experimental \citep{tran2013air} and numerical \citep{hendrix2016universal} results. It can be observed that the simulated bubble volumes are in quantitative agreement with the reported experimental measurements over a range of impact conditions. Moreover, the present results are consistent with the earlier experimental and numerical studies \citep{tran2013air,hendrix2016universal,Marston_2011,bouwhuis2012maximal,Duchemin2003}.

\section{Results and discussion}\label{sec:dis}

As discussed in the introduction, at low impact velocities, a droplet floats on the free surface, supported by an air cushion between the droplet and the pool, before merging into the pool \citep{Charles1960b, thoroddsen2000pof, Honey2006, blanchette2006partial}. Subsequently, the air film moves out, and the droplet makes contact with the free surface of the pool. This results in the formation of an opening at the contact point as the droplet merges with the liquid pool. The resultant high capillary pressure around this opening, caused by the large curvature, triggers rapid expansion of the neck, generating capillary waves that propagate along the droplet surface and the free surface of the pool. As the neck reaches its maximum radius, it begins to retract toward the axis of symmetry in an effort to minimize surface energy, thereby generating inward-directed horizontal momentum (negative horizontal momentum). Concurrently, the generated capillary waves converge at the tip of the droplet, exerting a vertical pull. The vertical pull and retraction of the neck result in the formation of a liquid column. Over time, the height of the column decreases as the droplet liquid continuously drains into the pool. The coalescence outcomes are primarily determined by the competition between vertical downward momentum and horizontal inward momentum \citep{blanchette2006partial,Ding2012,ray2010generation}. However, a recent study by \citet{Angeli2023} revealed that the Rayleigh-Plateau instability plays a crucial role in partial coalescence, with capillary waves only facilitating the formation of the liquid column. 

Previous studies have characterized the transition between coalescence outcomes using the Ohnesorge number ($\Oh$) and the Bond number ($\Bo$), reporting different critical Ohnesorge numbers, $\Oh_c$, associated with the transition between partial and complete coalescence \citep{Charles1960b,Aryafar2006,blanchette2006partial,blanchette2009dynamics,Gilet2007,ray2010generation,Behera2022}. As clarified in the introduction, this variability in the reported critical values arises primarily from differences in how $\Oh_c$ is defined and interpreted across studies. For instance, \citet{blanchette2006partial} reported a critical value of $\Oh_c = 0.018 \pm 0.002$ based on a combination of experiments and numerical simulations, whereas \citet{Aryafar2006} observed partial coalescence even for $\Oh_c > 1$, reflecting a unified definition based on the viscosity of the more viscous phase. Importantly, none of these values have been derived from theoretical foundations, underscoring the need for comprehensive investigations to establish more accurate values for $\Oh_c$. Additionally, these values were determined under conditions of negligible inertial forces. Recently,  \cite{behera2023investigation} examined the role of inertial forces in coalescence dynamics and presented a phase diagram of the critical Weber number ($\We_c$) as a function of $\Oh$ for a given $\Bo$. Similarly, \cite{anirudh2024coalescence} investigated the impact of $\We$ and drop shape on coalescence dynamics. However, we propose that the transitions between partial and complete coalescence regimes are governed by the interplay of all three non-dimensional parameters, namely Weber number ($\We$), Ohnesorge number ($\Oh$), and Bond number ($\Bo$). Furthermore, some researchers \citep{ray2010generation,Alhareth2020,Behera2022} have observed oscillations of the neck of the liquid column during the drainage of the liquid into the pool. However, the exact mechanism behind this phenomenon remains to be fully understood. Therefore, the present study aims to (i) investigate the effect of competitive effects of different governing forces, characterized by $\We$, $\Bo$, and $\Oh$, as a function of the shape of the drop on the coalescence dynamics, (ii) elucidate the mechanism behind the neck oscillations at the transition regime, which significantly influence the hydrodynamics of coalescence process, and (iii) generalize partial coalescence mechanisms.

%5
\begin{figure}
\centering
\includegraphics[width=0.95\textwidth]{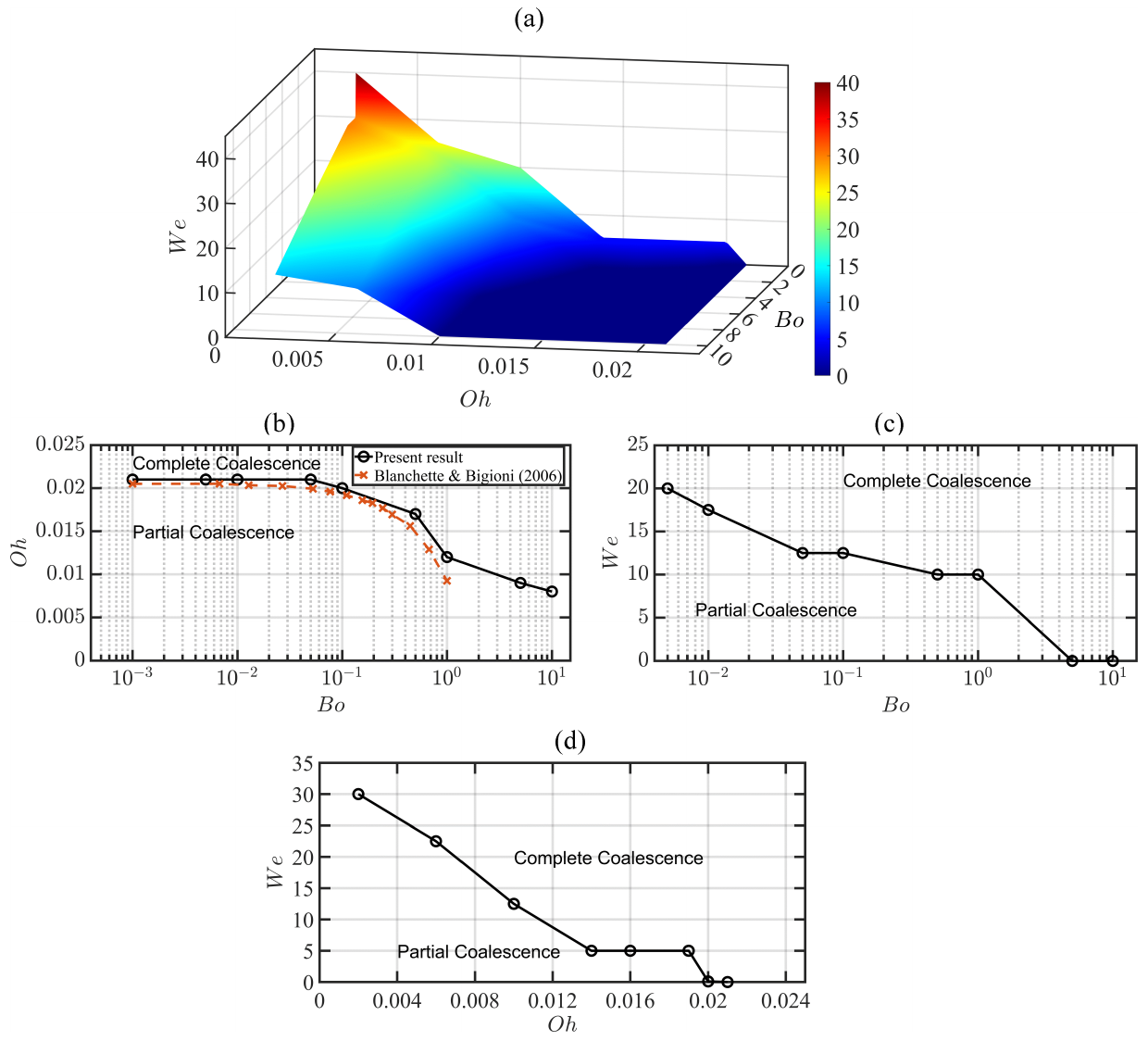} 
\caption{(a) Regime map showing the variation of the critical Weber number ($\We_c$) across the $\We–\Oh–\Bo$ parameter space, with the color bar representing the magnitude of $\We_c$. Cross-sectional phase diagrams highlight the boundaries between partial and complete coalescence regimes in (b) $\Oh–\Bo$ space at $\We = 0$, (c) $\We–\Bo$ space for $\Oh = 0.01$, and (d) $\Oh–\We$ space at $\Bo = 0.1$. Panel (b) also includes a comparison of our findings with the results reported by \citet{blanchette2006partial}.}
\label{fig:combined}
\end{figure}

\begin{figure}
\centering
\includegraphics[width=0.9\textwidth]{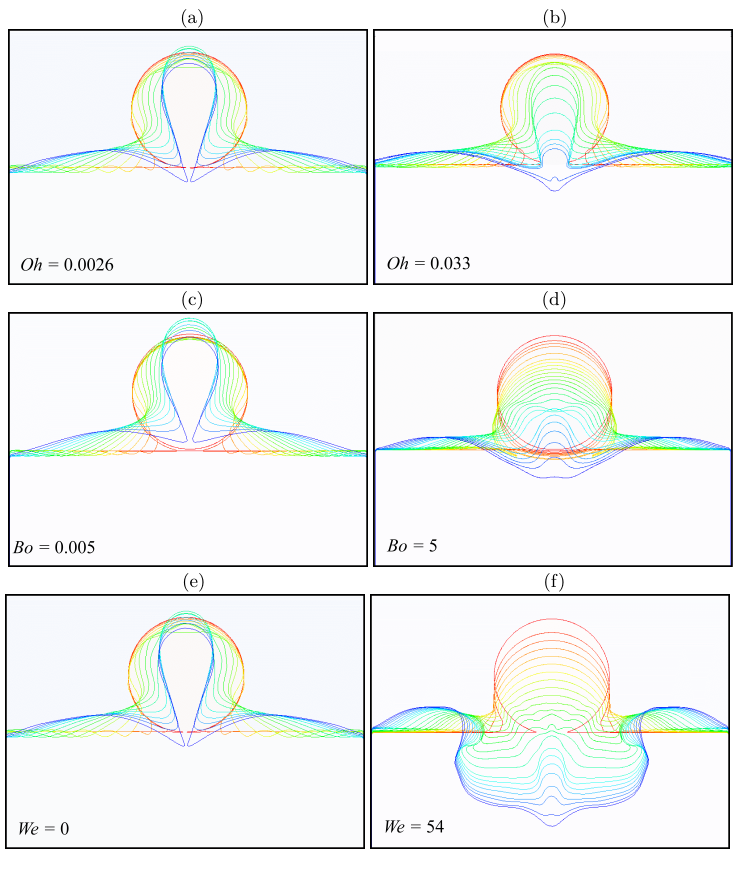}
\caption{Effect of (a–b) Ohnesorge number ($\Oh$) with $\AR = 1.0$, $\We = 0$, and $\Bo = 0.5$; (c–d) Bond number ($\Bo$) with $\AR = 1.0$, $\We = 0$, and $\Oh = 0.014$; and (e–f) Weber number ($\We$) with $\AR = 1.0$, $\Bo = 0.5$, and $\Oh = 0.0026$ on the coalescence dynamics of a liquid drop on a liquid pool. In each panel, time progression is indicated by a color change from red to blue, albeit with differing time scales across the panels.}
\label{fig_combinedb}
\end{figure}

A comprehensive investigation into coalescence dynamics has been conducted by systematically varying the fluid properties, impingement velocity, and drop diameter over a broad range of $\We$, $\Oh$, and $\Bo$. These dimensionless numbers quantify the relative contributions of inertial, viscous, capillary, and gravitational forces governing the coalescence process (Figure~\ref{fig:combined}). Since all simulations in the present study correspond to an air–liquid interface with negligible surrounding-fluid viscosity ($Oh_2 \ll 1$), the Ohnesorge number reported here corresponds to $Oh_1$, defined using the viscosity of the drop–pool liquid. Figure~\ref{fig:combined}(a) identifies the transition to complete coalescence by showing contours of the critical Weber number ($\We_c$) as a function of $\Oh$ and $\Bo$ for a spherical drop $(\AR=1)$. To better interpret this complex three-dimensional $\We$–$\Oh$–$\Bo$ regime map, it is useful to examine the pairwise interactions between these dimensionless parameters by holding one constant, as illustrated in Figures~\ref{fig:combined}(b–d). Figure~\ref{fig_combinedb}(a–f) demonstrates the effects of the individual parameters, namely $\We$, $\Oh$, and $\Bo$, on the coalescence dynamics by varying one dimensionless parameter at a time while keeping the other two fixed. Figure~\ref{fig:combined}(b) presents the critical Ohnesorge number ($\Oh_c$) as a function of $\Bo$ in the limit of negligible inertia ($\We \approx 0$). The results show that both increasing $\Oh$ and $\Bo$ promote complete coalescence. The specific effects of $\Oh$, illustrated in Figure~\ref{fig_combinedb}(a–b), indicate that viscous forces increasingly suppress capillary waves at higher values of $\Oh$. Capillary forces are essential for the formation and elongation of the rising liquid column. At a lower Ohnesorge number ($\Oh = 0.0026$), weaker viscous diffusion allows capillary-wave–induced perturbations to grow, resulting in an apex height that exceeds the initial drop diameter (Figure~\ref{fig_combinedb}a). In contrast, at a higher Ohnesorge number ($\Oh = 0.033$), enhanced viscous diffusion suppresses these perturbations, and the apex height remains below the initial drop diameter (Figure~\ref{fig_combinedb}b). Such suppression accelerates vertical collapse relative to horizontal collapse, thereby favoring complete coalescence. At sufficiently high $\Oh$, this mechanism enhances the drainage of liquid from the droplet, leading to complete coalescence. As evident from Figure~\ref{fig:combined}(b), the variation of the critical Ohnesorge number, $Oh_c$, with the Bond number, $Bo$, is weak at low $Bo$, indicating a negligible influence of gravitational forces. As $\Bo$ increases, this threshold $\Oh$ decreases, indicating a shift in dominance from capillarity to gravity. This trend is consistent with the observations of \citet{blanchette2006partial}, although minor discrepancies in the critical values may arise due to differences in drop initialization or air property modeling (Figure \ref{fig:combined}b). As illustrated in Figure~\ref{fig_combinedb}(c–d), the drainage of liquid from the drop into the pool increases with increasing $\Bo$, as gravitational forces increasingly dominate over capillary forces, leading to accelerated vertical collapse relative to horizontal collapse.

Figure~\ref{fig:combined}(c) depicts the variation of the critical Weber number, $\We_c$, with $\Bo$ at a fixed $\Oh = 0.01$. It is observed that $\We_c$ decreases as $\Bo$ increases. Focusing on the effect of $\We$, Figures~\ref{fig_combinedb}(e–f) illustrate that increasing $\We$ enhances inertial effects relative to capillarity, accelerating the drainage of the drop liquid into the pool and leading to a significant reduction in the apex height at a given time. This behavior favors the transition from partial to complete coalescence. As seen in Figure~\ref{fig_combinedb}(f), complete coalescence can occur even while upward-propagating capillary waves continue to distort the drop apex. This observation aligns with the conclusions of \cite{Gilet2007}, who reported that capillary wave convergence alone cannot explain partial coalescence. Since the vertical collapse rate increases relative to the horizontal collapse rate with increasing $\We$ and $\Bo$, the critical Weber number $\We_c$ decreases as $\Bo$ increases. Figure~\ref{fig:combined}(d) shows the variation of $\We_c$ with $\Oh$ at a fixed $\Bo = 0.1$. Since both $\We$ and $\Oh$ promote complete coalescence, $\We_c$ decreases as $\Oh$ increases. These two-dimensional regime maps (Figure~\ref{fig:combined}(b-d)) offer insights into the interplay of inertia, viscosity, and gravity in determining coalescence behavior. Figure~\ref{fig:combined}(a) synthesizes these insights by presenting a unified three-dimensional surface separating the regimes of partial and complete coalescence across the $\We$–$\Oh$–$\Bo$ space. For each pair of $\Oh$ and $\Bo$, a corresponding $\We_c$ marks the transition boundary. In the low-$\Oh$, low-$\Bo$ region, capillarity dominates, and $\We_c$ increases sharply. As $\Bo$ increases in this regime, $\We_c$ decreases due to enhanced gravitational drainage. With increasing $\Oh$, the system enters a viscous-dominated regime where the transition becomes largely independent of $\We$ and $\Bo$, as indicated by the plateau in the blue region of the figure. Beyond a threshold value of $\Oh$, the system predominantly exhibits complete coalescence. The present simulations estimate this critical Ohnesorge number to be $\Oh_c = 0.021$, above which complete coalescence occurs regardless of $\We$ and $\Bo$. This critical Ohnesorge number agrees with the threshold values reported by \citet{blanchette2006partial} and \citet{Gilet2007} under low-$\We$ conditions and negligible viscous effects of the surrounding fluid ($\Oh_2 \ll 1$). Previous studies have shown that low values of $\Bo$ and $\Oh$ generally lead to partial coalescence of the drop \citep{blanchette2006partial, Gilet2007, Aryafar2006}. However, the present study demonstrates that complete coalescence can still occur at low $\Bo$ and $\Oh$ provided that $\We_c \geq 40$ for $AR = 1$.

%6

 %7
\begin{figure}
\centering
\includegraphics[width=0.8\textwidth]{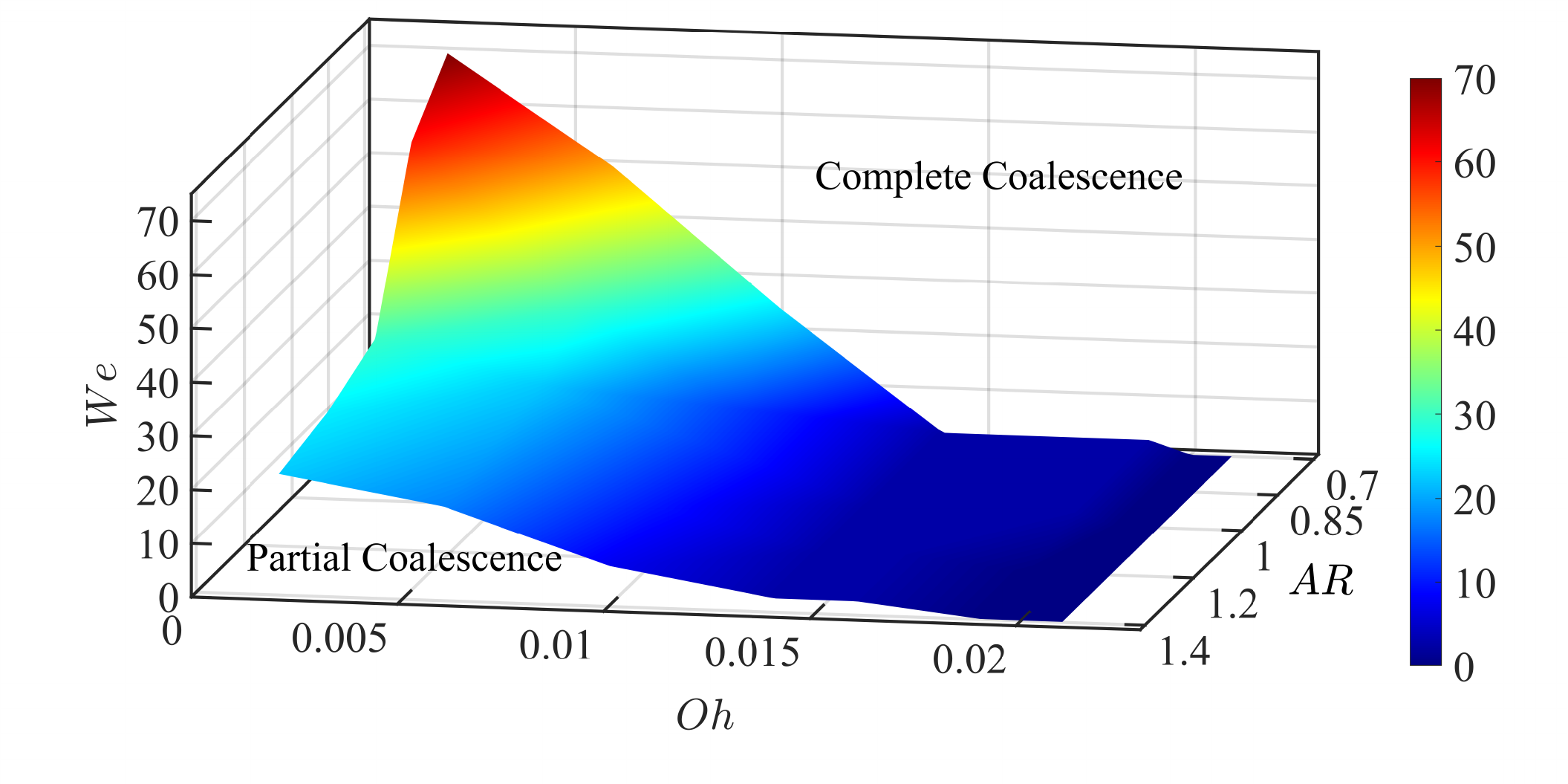}
\caption{Regime map illustrating the variation of the critical Weber number ($\We_c$), defined as the threshold Weber number separating partial and complete coalescence, across the $\We$–$\Oh$–$\AR$ parameter space. The color bar indicates the magnitude of $\We_c$. The regions below and above the surface of $We_c$ denote partial coalescence and complete coalescence, respectively.}
\label{fig_vdmap}
\end{figure}

%8
\begin{figure}
\centering
\includegraphics[width=0.85\textwidth]{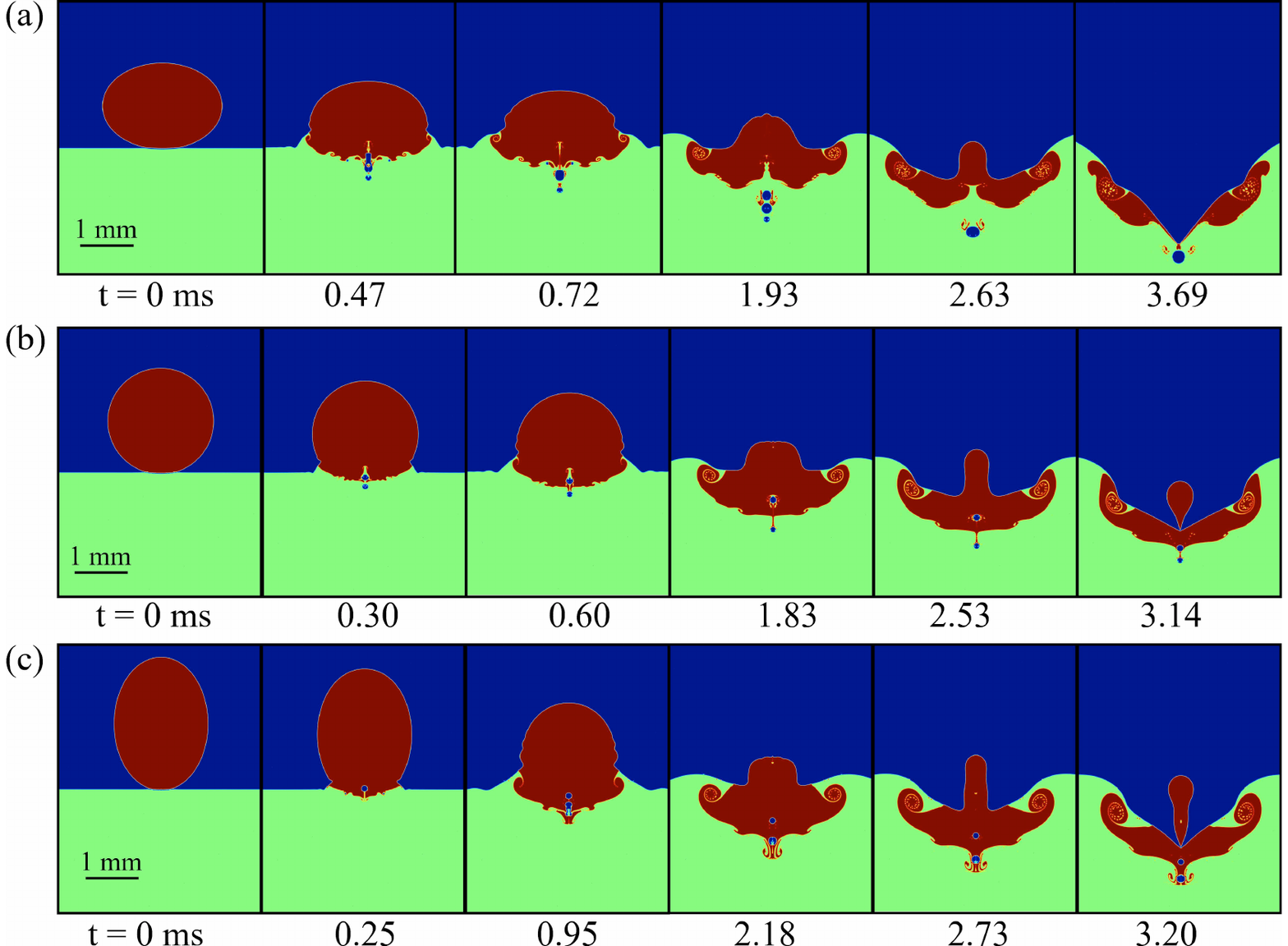}
\caption{Temporal evolution of coalescence dynamics for (a) an oblate drop ($\AR = 0.7$), (b) a spherical drop ($\AR = 1$), and (c) a prolate drop ($\AR = 1.4$). The rest of the parameters are $\We = 18$, $\Bo = 0.5$, and $\Oh = 0.0026 $.}
\label{fig_sf}
\end{figure}

%9
\begin{figure}
\centering
\includegraphics[width=0.85\textwidth]{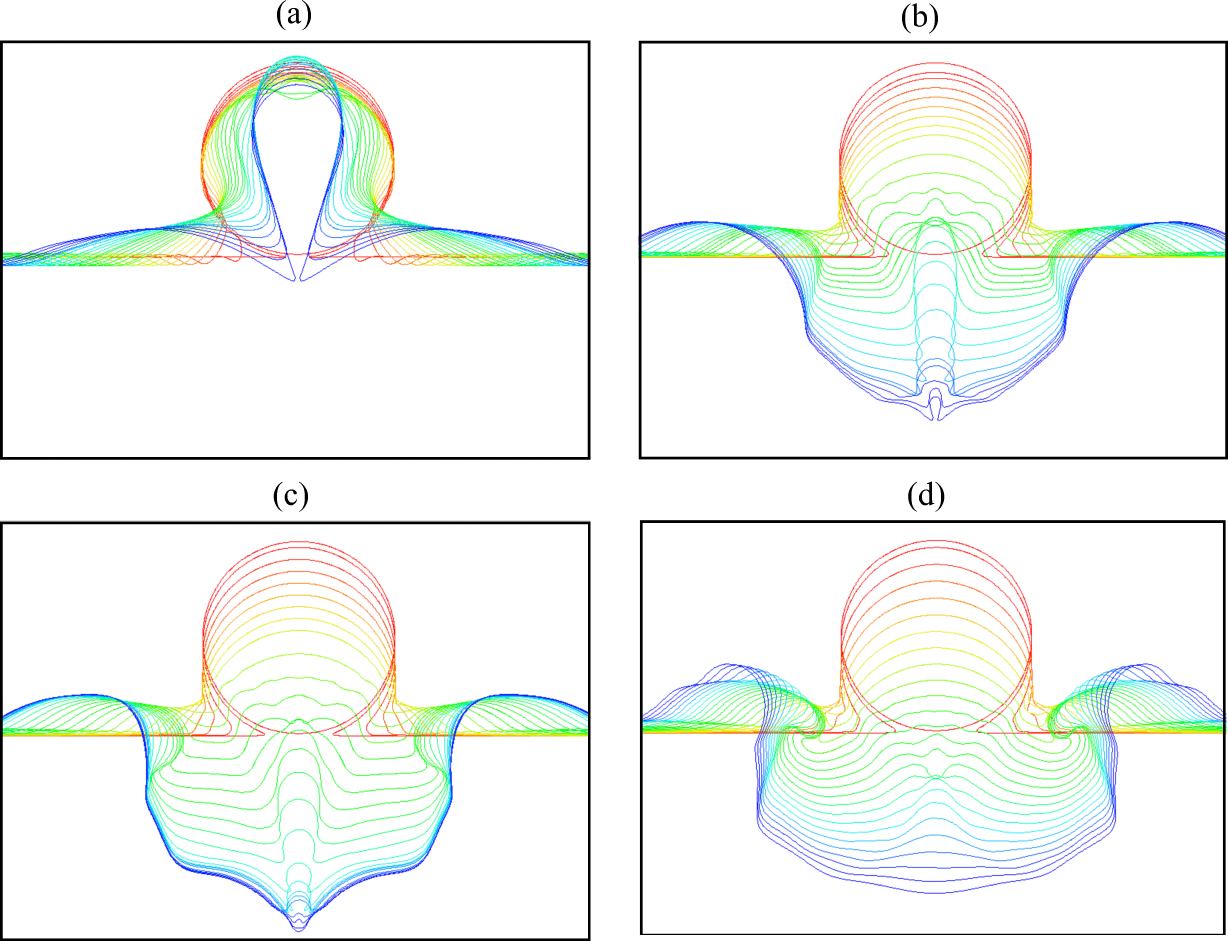}
\caption{Temporal evolution of the shape of the drop during coalescence in different regimes, categorized by the oscillations of the neck of the liquid column, for a primary drop with $\Bo = 0.5$, $\Oh = 0.0026$, and $\AR = 1.0$. (a) first-stage partial coalescence for $\We = 0$, (b) second-stage partial coalescence for $\We = 28$, (c) second-stage complete coalescence for $\We = 47$, and (d) first-stage complete coalescence for $\We = 90$.}
\label{fig_neck}
\end{figure}

The shape of a drop is another critical factor influencing the formation of secondary droplets. Figures \ref{fig_vdmap} and \ref{fig_sf} illustrate how drop shape affects the coalescence outcome on a liquid bath. Figure \ref{fig_vdmap} shows a regime map in the $\Oh$–$\We$ space, delineating the boundary between partial and complete coalescence for drops with different aspect ratios ($\AR$) at a fixed Bond number ($\Bo = 0.10$). It can be seen that, for a given $\Oh$, the critical Weber number ($\We_c$) decreases with decreasing aspect ratio, indicating that prolate-shaped drops ($\AR > 1$) are more prone to partial coalescence than spherical ($\AR = 1$) or oblate-shaped drops ($\AR < 1$). While the dependence of $\We_c$ on $\Oh$ was discussed earlier (see Figure \ref{fig:combined}), Figure \ref{fig_vdmap} highlights the additional influence of drop shape, showing that drops with higher $\AR$ require lower $\We$ to undergo partial coalescence. The temporal evolution of the coalescence process for oblate ($\AR = 0.7$), spherical ($\AR = 1$), and prolate-shaped drops ($\AR = 1.4$) with $\We = 18$, $\Bo = 0.5$, and $\Oh = 0.0026$ is shown in Figures \ref{fig_sf}(a–c), respectively. For drops with the same equivalent diameter ($D_{eq}$), the tendency to form secondary droplets is highest for prolate-shaped drops, followed by spherical, and lowest for oblate-shaped drops. This trend arises from the higher drainage rate of oblate-shaped drops due to their flatter base, which produces a larger neck opening (observed at $t = 0.47$ ms in Figure \ref{fig_sf}a). This larger neck enhances vertical downward momentum relative to horizontal inward momentum, promoting complete coalescence. In contrast, prolate-shaped drops have a narrower neck (observed at $t = 0.25$ ms in Figure \ref{fig_sf}c), which restricts drainage and delays vertical collapse, facilitating rapid pinch-off of a secondary droplet.
 
Figure~\ref{fig_neck}(a–d) illustrates the coalescence dynamics of a spherical primary drop with $\Bo = 0.5$ and $\Oh = 0.0026$ at varying Weber numbers selected to highlight distinct coalescence outcomes governed by neck oscillation behavior. While the partial and complete coalescence regimes have been extensively studied, the transitional dynamics between them have received comparatively less attention. Our investigation reveals that the transition from partial to complete coalescence is not abrupt; rather, it is mediated by neck oscillations, which can occur multiple times during the coalescence process. In Figures~\ref{fig_neck}(b–c), two distinct neck oscillation cycles are observed. Based on the number and outcome of these oscillations, we classify the coalescence process into four distinct regimes: (a) First-stage partial coalescence, where pinch-off occurs during the first oscillation; (b) Second-stage partial coalescence, where pinch-off occurs during the second oscillation; (c) Second-stage complete coalescence, where full merging occurs during the second oscillation; and (d) First-stage complete coalescence, where complete merging happens during the first oscillation. We observe that as the Weber number increases, the drop sequentially transitions through these regimes: from first-stage partial to second-stage partial, followed by second-stage complete, and eventually first-stage complete coalescence. While neck oscillations during coalescence on a liquid pool have been reported in previous studies \citep{ray2010generation, Alhareth2020, Behera2022}, the systematic identification and classification of all four coalescence regimes, as demonstrated in Figure~\ref{fig_neck}(a–d), provide a distinctive contribution of the present study.

%10
\begin{figure}
\centering
\includegraphics[width=0.3\textwidth]{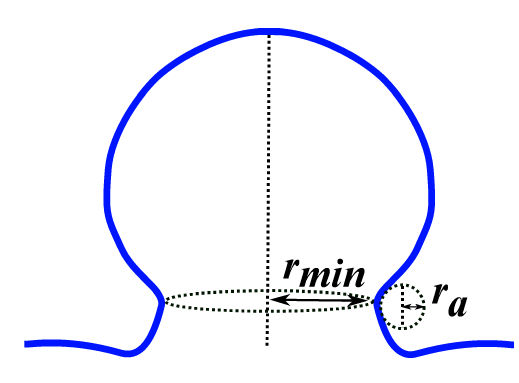}
\caption{Schematic diagram illustrating the azimuthal ($r_{\min}$) and axial ($r_{a}$) curvatures at the neck during the coalescence.}
\label{fig_curvature}
\end{figure}

To further understand the neck oscillation dynamics, it is essential to consider the roles of both azimuthal and axial curvatures at the neck, as schematically illustrated in Figure~\ref{fig_curvature}. In the transitional regime, the rates of horizontal and vertical collapse become comparable. As the cylindrical liquid column diminishes into a smaller droplet during the final stage of coalescence, the azimuthal and axial curvatures in the neck region play a critical role \citep{Alhareth2020, Zhang2009}. Specifically, the radius of curvature in the axial direction ($r_a$) becomes smaller than that in the azimuthal direction ($r_{\min}$). When $r_{\min} > r_a$, the capillary pressure, given by $\Delta{p} = \sigma (1/r_{\min} - 1/r_a)$, becomes negative, leading to an outward expansion of the neck. This expansion continues until $r_a$ exceeds $r_{\min}$, at which point the capillary pressure changes sign once again. During this retraction phase, the neck collapses inward and downward, forming a cylindrical liquid column. The final outcome, whether second-stage partial or second-stage complete coalescence, depends on the competition between the horizontal and vertical collapse rates.

%11
\begin{figure}
\centering
\includegraphics[width=0.95\textwidth]{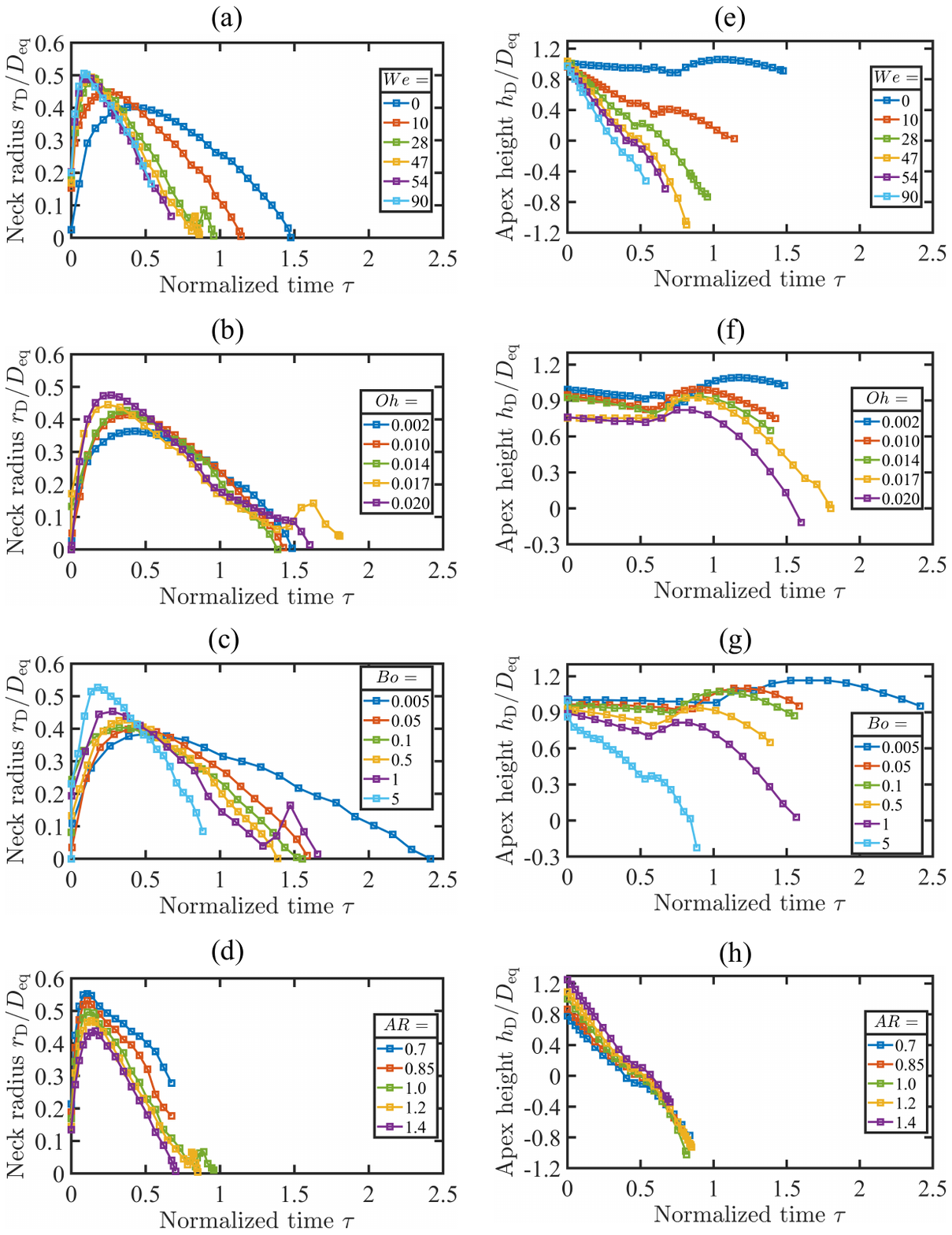}
\caption{Variations of the (a–d) normalized neck radius $(r_D/D_{eq})$ and (e–h) normalized apex height $(h_D/D_{eq})$ of the drop with normalized time $(\tau)$.
(a,e) Effect of Weber number ($\We$), with fixed $\AR=1.0$, $\Bo = 0.5$ and $\Oh = 0.0026$;
(b,f) effect of Ohnesorge number ($\Oh$), with $\AR=1.0$, $\We = 0$ and $\Bo = 0.5$;
(c,g) effect of Bond number ($\Bo$), with $\AR=1.0$, $\We = 0$ and $\Oh = 0.014$;
(d,h) effect of drop aspect ratio ($\AR$), with $\Bo = 0.5$, $\Oh = 0.0026$ and $\We = 23$.}
\label{fig_neckg}
\end{figure}

The temporal evolution of the retracting neck and the apex height of the drop, which are associated with the horizontal and vertical collapse rates, respectively, governs the outcome of the coalescence process. To characterize this hydrodynamic behavior, we measured the time-dependent variations of the neck radius and the apex height. The transition phase between partial and complete coalescence is characterized by oscillations in the neck dynamics, which serve as a clear indicator of this intermediate regime. Figures~\ref{fig_neckg}(a–d) show the normalized neck radius ($r_D/D_{eq}$) plotted against the normalized time ($\tau$), highlighting the effects of the Weber number ($\We$), Ohnesorge number ($\Oh$), Bond number ($\Bo$), and aspect ratio of the primary drop ($\AR$). In each case, one dimensionless parameter is varied while the others are held constant. Across all cases, the neck radius initially increases due to the excess capillary pressure within the drop. Once it reaches a maximum, the radius subsequently decreases as the system evolves to minimize its surface energy. Figure~\ref{fig_neckg}(a) depicts the effect of $\We$ on the temporal evolution of the neck radius at fixed $\Bo$ and $\Oh$. It can be seen that as $\We$ increases, the neck velocity during both expansion and retraction increases, indicating a growing dominance of inertial forces over capillary forces. Although a larger $\We$ leads to a greater maximum neck radius, which might intuitively slow retraction, this trend instead reflects a complex interplay between inertia and surface tension. Figure~\ref{fig_neckg}(b) displays the downward motion of the drop via the evolution of the apex height for different $\We$. Initially, the apex height decreases approximately linearly. Subsequently, a distinct mound forms due to upward-traveling capillary waves generated during coalescence. As $\We$ increases, the height of this mound decreases, and the initial linear stage shortens, indicating that inertial forces increasingly suppress capillary effects. The resulting enhancement in vertical collapse rate plays a key role in driving the transition from partial to complete coalescence. This is evident from Figure~\ref{fig_neckg}(a), where the neck exhibits a single oscillation for both first-stage partial ($\We = 0$, 10) and first-stage complete coalescence ($\We = 54$, 90), while it oscillates twice during second-stage partial ($\We = 28$) and second-stage complete coalescence ($\We = 47$).

The effect of viscosity of the drop characterized by the Ohnesorge number ($\Oh$) is examined in Figure~\ref{fig_neckg}(b,f). As $\Oh$ increases, viscous damping becomes more prominent, suppressing capillary waves. Interestingly, the neck retraction velocity and the maximum neck radius increase with increasing $\Oh$. This behavior is evident in the apex height evolution (Figure~\ref{fig_neckg}b), where increasing viscosity diminishes the height of the capillary-induced mound, facilitating a faster vertical collapse and promoting the transition from partial to complete coalescence. A single neck oscillation is observed for both first-stage partial ($\Oh = 0.002$, 0.010) and first-stage complete coalescence ($\Oh = 0.020$), while two oscillations occur in the case of second-stage complete coalescence ($\Oh = 0.017$).

The influence of gravity, represented by the Bond number ($\Bo$), is shown in Figures~\ref{fig_neckg}(c,f). As $\Bo$ increases, gravitational forces become stronger than surface tension, accelerating neck retraction and increasing the maximum neck radius. A higher $\Bo$ also enhances liquid drainage from the drop, reducing both the stagnation period and mound height, as evident in Figure~\ref{fig_neckg}(f). This results in a faster vertical collapse and facilitates the transition from first-stage partial ($\Bo = 0.005$, 0.05, 0.1, 0.5) to first-stage complete coalescence ($\Bo = 5$), via second-stage complete coalescence ($\Bo = 1$). Figure~\ref{fig_neckg}(d,h) examines the effect of drop shape, characterized by the aspect ratio ($\AR$). The neck opens more widely for prolate drops ($\AR > 1$), resulting in a larger neck radius at any given time compared to spherical or oblate drops. A lower $\AR$ corresponds to reduced neck velocity and a delayed horizontal collapse. Additionally, drop shape influences apex height due to geometric constraints: oblate drops exhibit lower apex heights, promoting faster vertical collapse. As the horizontal collapse rate decreases and the vertical collapse rate increases when transitioning from prolate to oblate shapes via spherical drops, the coalescence behavior shifts from first-stage partial ($\AR = 1.4$) to first-stage complete coalescence ($\AR = 0.7$, 0.85), passing through the transition phase with double neck oscillations ($\AR = 1$, 1.2), characteristic of second-stage partial coalescence.

%12
\begin{figure}
\centering
\includegraphics[width=0.9\textwidth]{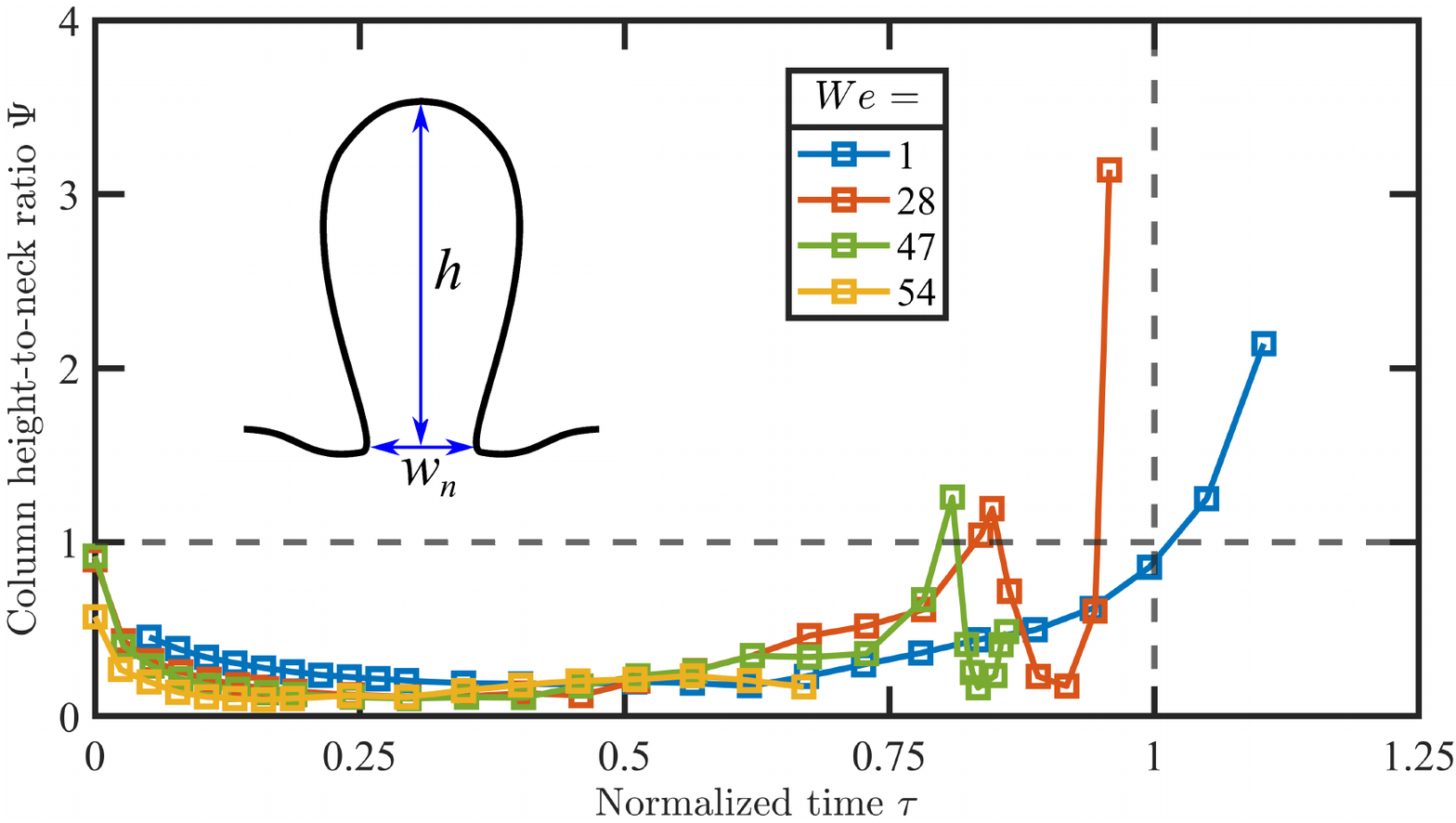}
\caption{Temporal variation of $\psi=h/\pi w_n$ for different Weber numbers. Here, $\We = $ 1, 28, 47, and 54 correspond to the first-stage partial coalescence, second-stage partial coalescence, second-stage partial coalescence and first-stage complete coalescence scenarios. The rest of the parameters are $\Bo = 0.5$, $\Oh = 0.0026 $ and $\AR = 1$.}
\label{fig_cylinder}
\end{figure}

To examine the potential role of Rayleigh–Plateau instability in the pinch-off process, Figure~\ref{fig_cylinder} shows the temporal evolution of the column height-to-neck ratio, $\psi = h/\pi w_n$, for various Weber numbers ($\We$). The rest of the parameters are $\Bo = 0.5$, $\Oh = 0.0026 $, and $\AR = 1$.  Here, $h$ represents the height of the liquid column, and $w_n$ is the neck diameter, both measured following the methodology of \citet{Angeli2023}. In Figure~\ref{fig_cylinder}, the selected cases, $\We = 1$, 28, 47, and 54, correspond to first-stage partial coalescence, second-stage partial coalescence, second-stage complete coalescence, and first-stage complete coalescence, respectively. For the complete coalescence case ($\We = 54$), $\psi$ remains consistently below one throughout the pinch-off process. In contrast, for first-stage partial coalescence ($\We = 1$), $\psi$ exceeds one just prior to the pinch-off of the secondary droplet. This indicates that if the Rayleigh–Plateau instability were responsible for the pinch-off, the instability would have been expected to manifest much earlier in the pinch-off process.  Similarly, in second-stage coalescence cases ($\We = 28$ and 47), $\psi$ temporarily exceeds one during the neck retraction phase but decreases without leading to a pinch-off. These findings indicate that Rayleigh–Plateau instability alone cannot fully explain the coalescence dynamics observed for the parameters considered in the present study.

%13
\begin{figure}
\centering
\includegraphics[width=0.95\textwidth]{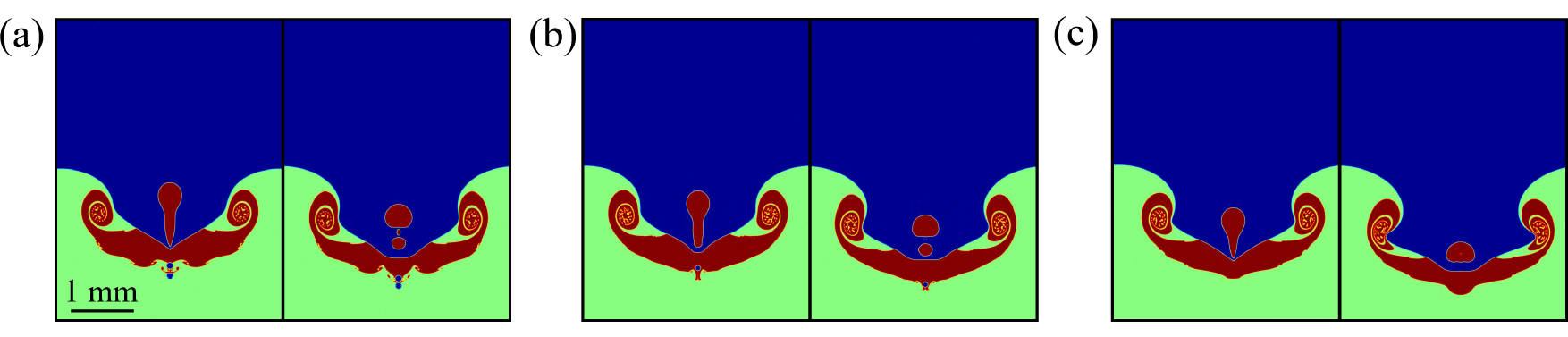}
\caption{Formation of multiple secondary droplets for a prolate-shape primary drop with $\Bo = 0.5$, $\Oh = 0.0026 $, and $\AR = 1.4$ impacting a liquid pool with different Weber numbers. (a) $\We = 28$ (three secondary droplets), (b) $\We = 33$ (two secondary droplets) and (c) $\We = 40$ (one secondary droplet).}
\label{fig_multipledrops}
\end{figure}

Finally, we explore the dynamics underlying the formation of multiple secondary drops, as shown in Figure \ref{fig_multipledrops}. This phenomenon was initially reported by \cite{Charles1960b} for the coalescence of larger water drops ($D_{eq} > 6$ mm) at a benzene / water interface, attributed to the Rayleigh capillary instability. \citet{Angeli2023} characterized the probability of the formation of multiple droplets by measuring the height to circumference ratio ($\psi$) of the deformed drop in the column formation stage due to capillary pull. Recently, through three-dimensional numerical simulations, \cite{anirudh2024coalescence} also observed the formation of multiple secondary drops during the coalescence of an oblate drop with a large aspect ratio. In contrast, in the present work, we observe that the generation of multiple secondary droplets occurs when the shape of the drop at the time of contact is prolate (Figure \ref{fig_multipledrops}). As depicted in the figure, the pinching off of secondary droplets occurs first, followed by their fragmentation into multiple droplets. This suggests that when the primary drop is pinched off because of negative horizontal momentum, the varicose capillary waves generated during the pinching-off process tend to break this secondary droplet because of the Rayleigh-Plateau instability. When a prolate drop comes into contact with a pool, a secondary drop forms as an extended liquid column, compared to an oblate or spherical droplet shape. This extended column assists the Rayleigh-Plateau instability in breaking into multiple droplets. It is worth emphasizing that the Rayleigh–Plateau instability typically contributes to interfaces that are locally cylindrical, a condition not always satisfied during partial coalescence. When the falling drop contacts the liquid pool, capillary waves travel from the point of contact toward the apex, generating an upward pull. During the same time, part of the liquid in the drop drains downward due to vertical momentum, while strong horizontal momentum inward at the neck initiates pinching, resulting in the formation of a secondary droplet. If capillary-induced upward motion creates a columnar structure ($\psi > 1$) with a short cylindrical region of nearly constant radius, the Rayleigh-Plateau instability can be activated. Additionally, following the formation of the secondary droplet, varicose perturbations on its elongated, prolate shape can give rise to this instability, potentially leading to its further breakup. Our findings reveal that increases in the Weber, Ohnesorge, and Bond numbers enhance liquid drainage into the pool, yielding smaller secondary droplets and only slightly elongating the liquid column. As a result, the likelihood of droplet breakup due to Rayleigh–Plateau instability diminishes.

\section{Concluding remarks}\label{sec:conc}

We investigate the dynamics of liquid drop coalescence in a deep liquid pool through axisymmetric numerical simulations using Gerris \citep{Popinet2003,Popinet2009}, an open-source finite volume flow solver. This study aims to generalize the mechanisms of partial coalescence for a range of droplet sizes, shapes, impact velocities, and liquid properties and elucidate the role of neck oscillations, which significantly influence the subsequent coalescence process and the formation of secondary droplets. The effects of relevant dimensionless parameters, such as Weber $(\We)$, Ohnesorge $(\Oh)$, Bond $(\Bo)$ numbers, and the aspect ratio $(\AR)$ of the drop are also examined. By performing extensive simulations, we construct a three-dimensional regime map in the $\We–\Oh–\Bo$ space that delineates the transition boundaries between partial and complete coalescence. Our results reveal four distinct coalescence regimes based on the oscillatory behavior of the neck of the liquid column, namely (i) first-stage partial coalescence, where pinch-off occurs during the initial neck oscillation; (ii) second-stage partial coalescence, with pinch-off occurring during the second oscillation; (iii) first-stage complete coalescence, characterized by full merging during the first oscillation; and (iv) second-stage complete coalescence, in which complete merging occurs during the second oscillation. The temporal evolution of the neck radius and apex height shows that the transition between the two regimes of coalescence, namely the partial and complete coalescence is driven by the competition between horizontal and vertical collapse dynamics, which is further influenced by inertia, viscosity, gravity, and the morphology of the primary droplet. Neck oscillations emerge as a robust indicator of the coalescence regime, with the number of oscillation cycles strongly correlating with the stage of coalescence across a wide range of $\We$, $\Oh$, $\Bo$, and $\AR$ values. Furthermore, our analysis of the height-to-neck ratio of the liquid column suggests that the Rayleigh–Plateau instability does not adequately explain the partial coalescence behavior observed under the investigated conditions. Overall, this study offers a comprehensive understanding of the hydrodynamic mechanisms governing distinct coalescence dynamics over a wide range of parameters.
\\

\section*{ Declaration of Interests:} The authors report no conflict of interest.

\section*{Acknowledgements:} G.B. acknowledges his gratitude to J. C. Bose National Fellowship of SERB, Government of India (grant: JBR/2020/000042). K.C.S. thanks IIT Hyderabad for the financial support through grant IITH/CHE/F011/SOCH1. M.R.B. acknowledges financial support from BITS-Pilani, K. K. Birla Goa Campus through Grant No. NFSG/GOA/2024/G0978 and support from IIT Kanpur under the Institute PDF.

%\bibliographystyle{jfm}
%\bibliography{bibl.bib}

\end{document}